\documentclass[pra,twocolumn,aps,floatfix,showpacs,tightenlines,superscriptaddress,amsmath,amssymb]{revtex4}
\usepackage{amssymb,amsmath,amsthm,amsbsy,epsfig,color,graphicx,times}
\usepackage[ansinew]{inputenc}
\usepackage[english]{babel}
\usepackage{accents}
\usepackage{color}
\usepackage{braket}
\usepackage{amsfonts}
\usepackage{booktabs}
\usepackage{tabularx}
\usepackage{array}
\usepackage{hyperref}
\hypersetup{pdfstartview=FitH}
\usepackage{latexsym}
\usepackage{multirow}
\usepackage{longtable}

\begin{document}

\title{Quench of a symmetry broken ground state}

\author{S. M. Giampaolo}
\affiliation{International Institute of Physics, UFRN, Anel Vi\'ario da UFRN, Lagoa Nova, 59078-970 Natal - RN, Brazil}
\author{G. Zonzo}
\affiliation{Dipartimento di Fisica ``E. R. Caianiello'', Universit\`a degli Studi di Salerno,
Via Ponte don Melillo 1, I-84084 Fisciano (SA), Italy}

\begin{abstract}
We analyze the problem of how different ground states associated to the same set of the Hamiltonian parameters evolve after a sudden quench. 
To realize our analysis we define a quantitative approach to the local distinguishability between different ground states of a magnetically 
ordered phase in terms of the trace distance between the reduced density matrices obtained projecting two ground states in the same subset.
Before the quench, regardless the particular choice of the subset, any system in a magnetically ordered phase is characterized by ground states 
that are locally distinguishable.
On the other hand, after the quench, the maximum of the distinguishability shows an exponential decay in time.  
Hence, in the limit of very large time, all the informations about the particular initial ground state are lost even if the systems are integrable.
We prove our claims in the framework of the magnetically ordered phases that characterize both the $XY$ model and $N$-cluster Ising models.
The fact that we find similar behavior in models within different classes of symmetry makes us confident about the generality of our results.
\end{abstract}

\maketitle

\section{Introduction}

Recent developments in the experimental control of systems realized with ultra cold atoms in optical 
lattices~\cite{Greiner2002,Kinoshita2006,Bloch2008} have disclosed the possible to test the predictions of the eigenstate thermalization 
hypothesis~\cite{Deutsch1991,Srednicki1994}.
In the framework of this theory, said $U$ a bipartite system $U=S\otimes T$, the subset $S$ can be seen as an independent system interacting with the 
thermal bath $T$. 
If the system $U$ is prepared in a state that is far from equilibrium, the time evolution of local quantities, i.e. quantities associated to operators 
with supports completely included in $S$, is indistinguishable by the time evolution of a system going towards a thermal equilibrium with a bath.
In other words a closed quantum system may locally thermalize~\cite{Rigol2008,Popescu2006,Polkovnikov2011,Linden2009,Gogolin2016}.
This implies that, in the steady state, local physical quantities will lose all the informations about the initial state with the exception of the 
effective temperature~\cite{Rigol2008,Rigol2012,Hortikar1997}. 

There are several ways to prepare a system far from equilibrium.
An important one is associated with the sudden quench of the Hamiltonian parameters.
The system, initially, prepared in an equilibrium state, undergoes to an abrupt change of the Hamiltonian parameters.
As a consequence, the system, in general, will be no more at the equilibrium and hence it will start to evolve under the action of the new 
Hamiltonian~\cite{Gritsev2007,Lamacraft2007,Rossini2009,Rossini2010,Canova2014,Zeng2015}.
For several models and initial conditions, under the effect of a quench, all the local physical quantities equilibrate exponentially in time and, 
at the end, the time evolution would produce a steady state that looks locally thermal~\cite{Barthel2008}.
However not all models are subjected to such thermalization process when placed away from the equilibrium. 
The discovery that the integrability of a model may allow it to avoid the thermalization process, triggers an intense discussion about the general 
relation between quantum integrability and thermalization in the long-time dynamics of strongly interacting complex quantum 
systems~\cite{Igloi2000,Calabrese2006,Cazalilla2006,Rigol2007,Barthel2008,Rigol2008,Eckstein2008}.

However the initial ground state may depend not only on the Hamiltonian parameters. 
In the presence of a degeneracy, for each single set of the parameters of the system, there will be an infinite number of ground states that may, or 
may not, differ each other for the expectation value of some local physical observables. 
In the case in which there is at least one single observable, with a finite support, for which the expectation value depends on the ground states, 
they are said to be locally distinguishable respect to all the subsets that include the support of the observable.
The most natural example of a system with locally distinguishable ground states is represented by a system in a magnetically ordered 
phase~\cite{Sachdev2000}. 
Independently on the subset taken into account, the different ground states can be characterized looking to the spin operator associated to the order 
parameter. 
Distinguishable ground states have very different physical properties.  
For example, only some specific ground states, i.e. the ones that maximize the order parameter, present a complete 
factorization~\cite{Rossignoli2008,Giampaolofatt} and a vanishing mutual information between very far spins~\cite{Hamma2016}. 
On the other hand, the symmetric ground states, i.e. the ground states for which the order parameter is zero, are the ones for which both the 
concurrence below the factorization point~\cite{Osterloh2006,Cianciaruso2014} and the Von Neumann entropy~\cite{Oliveira2008} 
reach their maximum values.

It is therefore natural to wonder if, in the integrable models, the distinguishability between the different ground states of a magnetically ordered 
phase is preserved after a sudden change of one or more parameters of the Hamiltonian. 
Because of the lack of thermalization of the integrable system, one can be driven to think that, in such cases, the steady state, obtained for very 
long time after the quench, will continue to preserve memory of the particular initial ground state. 
However, as we will see, this is not true. 
As we have just said, the lack of thermalization does not mean that all physical quantities do not thermalize, but only that there are some physical 
quantities, at least one, for which the thermalization process fails. 
Therefore if the distinguishability between the different ground states is associated to quantities for which the thermalization process works,
it can vanish also in integrable models.
In the present article we prove, for several integrable models that fall in different classes of symmetry, that the local distinguishability between 
different ground states in magnetically ordered phase disappears exponentially in time as result of the effect of a sudden quench of the Hamiltonian 
parameters.
To prove this result, in the next section we introduce a way to quantify the local distinguishability between the different ground states of the 
system, based on the trace distance between the reduced density matrices obtained from two different ground states projected in the same subset.
Thanks to such approach we show, in a very general way, that the local distinguishability between two ground states reaches the maximum if the two 
ground states are respectively: 1) one of the maximally symmetry broken ground state, i.e. one of the ground state for which the order parameter 
reaches the maximum; 2) one of the symmetric ground state that is one of the ground state that is also eigenstate of the parity operator.
We can then use these results to study the distinguishability, and its evolution after a quench, between the different ground states in several 
integrable models. 
In the Sec.III we study the effect of a quantum quench in the one-dimensional $XY$ model in the orthogonal external fields~\cite{Lieb1961,Barouch1970} 
while in Sec IV we focus our attention on the one-dimensional $N$-Cluster Ising models~\cite{Smacchia2011,Giampaolo2014,Giampaolo2015}.
Independently on the model, we show that, after a short transient, the local distinguishability disappears exponentially in time.
Consequently in the steady state that can be obtained for diverging time, all the informations about the particular ground state are completely lost. 
At the end, in the last section, we draw our conclusions.


\section{A quantitative approach to the distinguishability}

In this section we provide a quantitative approach to the distinguishability between two different ground states in the ferromagnetic phase.
To begin let us fix some points. 
All along the paper we will consider a one dimensional spin-$1/2$ system which dynamic is described by a translational invariant Hamiltonian 
$H_{\{\lambda\}}$ that depends on a set of parameters $\{\lambda\}$.
We assume that, regardless the values of the Hamiltonian parameters, $H_{\{\lambda\}}$ satisfies the parity symmetry respect to a spin directions
that for sake of simplicity and without loosing any generality, we fix to $z$.
This means that $H_{\{\lambda\}}$ commutes with the parity operator $P_z=\bigotimes_{i=1}^N\sigma_i^z$ where $N$ is the total number of spins in 
the system and $\sigma_i^z$ is the $z$ Pauli operator.
Because $[H_{\{\lambda\}},P_z]=0$ we have that the Hamiltonian and the parity operator admit a complete set of eigenstates in common. 
However, in the case in which the Hamiltonian shows degenerated eigenvalues, we may have eigenstates of the Hamiltonian that are not eigenstates of 
the parity. 
When this happens at the level of the ground state, we have the phenomenon known as a spontaneous symmetry breaking of which the magnetically ordered 
phase is the most famous example.

In the magnetically ordered phases of a one dimensional system made by spin-$1/2$, the Hamiltonian admits a twofold degenerated ground 
states~\cite{Rossignoli2008,Sachdev2000}. 
Among all the others, there will be two ground states that are also eigenstates of the parity with opposite eigenvalues. 
These two symmetric ground states, usually named the even $\ket{e_{\{\lambda\}}}$ ($P_z \ket{e_{\{\lambda\}}}=\ket{e_{\{\lambda\}}}$) and the odd 
$\ket{o_{\{\lambda\}}}$ ($P_z\ket{o_{\{\lambda\}}}=- \ket{o_{\{\lambda\}}}$) ground states, form a complete orthonormal base for the space made by 
all the ground states of $H_{\{\lambda\}}$.
Therefore all the ground states of $H_{\{\lambda\}}$ can be written in the form
\begin{equation}
 \label{groundstate}
 \ket{g_{\{\lambda\}}(u,v)} = u \ket{e_{\{\lambda\}}}+ v \ket{o_{\{\lambda\}}} \; ,
\end{equation}
where $u$ and $v$ are complex superposition amplitudes constrained by the normalization condition $|u|^2+|v|^2=1$. 

Because we are interesting on the local distinguishability between the different ground states, let us introduce a generic subset $S$ made by $l$ 
spins ($S=\{i_1, \cdots, i_l\}$).
The projection of the state $\ket{g_{\{\lambda\}}(u,v)} $ into $S$ is represented by the reduced density matrix, $\rho_{\{\lambda\}}(u,v,S)$, that is 
obtained tracing out all the degrees of freedom that fall outside $S$.
The reduced density matrix $\rho_{\{\lambda\}}(u,v,S)$ can be expressed in terms of the $l$-points spin correlation functions~\cite{Osborne2002} as
\begin{equation}
\label{reduced1}
\!\!\!  \rho_{\{\lambda\}}(u,v,S)\! = \!\frac{1}{2^l} \! \sum_{{\{\mu_i\}}} \!
 \bra{g_{\{\lambda\}}\!(u,v)} \! \hat{O}^{\{\!\mu_i\!\}}_S \! \ket{g_{\{\lambda\}}\!(u,v)}\!
   \hat{O}^{\{\!\mu_i\!\}}_S.
\end{equation}
In the above equation $\hat{O}^{\{\!\mu_i\!\}}_S=\sigma_{i_1}^{\mu_1}\otimes \sigma_{i_2}^{\mu_2}\otimes \ldots  \otimes \!\sigma_{i_l}^{\mu_l} $ is 
the tensor product of Pauli operators defined on the spins in $S$, $\{\mu_i\}$ is a set of $l$ variables where any single element range across 
\mbox{$\mu_i=0, \, x, \, y,\, z$}, the sum runs on all possible $\{\mu_i\}$ and $\sigma_i^0$ stands for the identity operator on the $i$-th spin.

With respect to the parity operator $P_z=\bigotimes_{i=1}^N\sigma_i^z$ any operator $\hat{O}^{\{\!\mu_i\!\}}_S$ can be classified in two different 
families.
The first is made by the operators $\hat{O}^{\{\!\mu_i\!\}}_S$ that commute with $P_z$ while the second is made by the operators that anti-commute 
with it and that bring even states in odd ones and viceversa. 
This fact plays a fundamental role when we try to evaluate the expectation value of the operator $\hat{O}^{\{\!\mu_i\!\}}_S$ on the symmetric
ground states $\ket{e_{\{\gamma\}}}$ and $\ket{o_{\{\gamma\}}}$.
In fact, because any operator $\hat{O}^{\{\!\mu_i\!\}}_S$ that anti-commutes with $P_z$ drives even states in odd ones, its expectation value on a
symmetric ground states have to be zero.
Not only.
It is well known that, in the thermodynamic limit, the expectation value of an operator that commutes with the parity is the same if evaluated on  
$\ket{e_{\{\gamma\}}}$ or on $\ket{o_{\{\gamma\}}}$~\cite{Sachdev2000,Rossignoli2008,Cianciaruso2014}.
We indicate such expectation value on one of the symmetric ground states as $\langle\hat{O}^{\{\mu_i\}}_S\rangle$.
As a consequence we have that the two symmetric ground states are always locally indistinguishable.  

Collecting together all the considerations made till now we have that the reduced density matrix $\rho_{\{\lambda\}}(u,v,S)$ in eq.~(\ref{reduced1}) 
can be rewritten as
\begin{equation}
\label{reduced2}
\rho_{\{\lambda\}}(u,v,S)  = {\rho}_{\{\lambda\}}^{sym}(S) + \chi_{\{\lambda\}}(u,v,S) \;.
\end{equation}
The density matrix $\rho^{sym}_{\{\lambda\}}(S)$ is obtained projecting one of the two symmetric ground state into $S$, and it is equal to
\begin{equation}
\label{rhotilde}
\rho^{sym}_{\{\lambda\}}(S)=\frac{1}{2^l} \sum_{{\{\mu_i\}}} 
 \langle \hat{O}^{\{\!\mu_i\!\}}_{S} \rangle
   \hat{O}^{\{\!\mu_i\!\}}_S\;,
\end{equation}
where the sum extends over all the operators $\hat{O}_S^{\{\mu_i\}}$ that commute with $P_z$.
On the other hand $\chi_{\{\lambda\}}(u,v,S)$ is an Hermitian traceless matrix that depends on the superposition parameters and is made by the 
contributions of all the operators $\hat{O}_S^{\{\mu_i\}}$ that anti-commutes with $P_z$.

Let us now introduce, for a generic spin operator $\hat{O}^{\{\mu_i\}}_S$ defined on $S$ and that anti-commutes with $P_z$, the operator 
$\hat{W}^{\{\mu_i\}}_{S\cup S+R}=\hat{O}^{\{\mu_i\}}_S \otimes \hat{O}^{\{\mu_i\}}_{S+R}$. Here $S+R$ is a new subset of spins of the system obtained 
from $S$ by a rigid spatial translation of $R$ and 
$\hat{O}^{\{\mu_i\}}_{S+R}=\sigma_{i_1+R}^{\mu_1}\! \otimes\! \sigma_{i_2+R}^{\mu_2}\! \otimes \!\ldots\!  \otimes \!\sigma_{i_l+R}^{\mu_l} $. 
Because both $\hat{O}^{\{\mu_i\}}_{S}$ and $\hat{O}^{\{\mu_i\}}_{S+R}$ anti-commutes with $P_z$, $\hat{W}^{\{\mu_i\}}_{S\cup S+R}$ will commute with 
the parity and hence its expectation value on a symmetric ground state can be different from zero. 
The expectation value of $\hat{O}^{\{\mu_i\}}_S$ on the maximum symmetry breaking state, obtained taking $u=\pm v=1/\sqrt{2}$, and hence the 
correlation function associated to such operator, is recovered exploiting the property of asymptotic factorization of products of local operators 
at infinite separation that yields to
\begin{equation}
\label{expected}
\langle \hat{O}^{\{\!\mu_i\!\}}_{S} \rangle =\sqrt{
\lim_{R \rightarrow \infty}  \langle \hat{W}^{\{\mu_i\}}_{S\cup S+R} \rangle} \;.
 \end{equation}

Starting from this state independent expression of the $\langle \hat{O}^{\{\!\mu_i\!\}}_{S} \rangle$ for operators that anti-commutes with the parity 
we obtain that eq.~(\ref{reduced2}) can be written as 
\begin{equation}
\label{reduced3}
\rho_{\{\lambda\}}(u,v,S)  = \rho^{sym}_{\{\lambda\}}(S) + (u^*v+v^*u)\tilde{\chi}_{\{\lambda\}}(S)  \;,
\end{equation}
where 
\begin{equation}
\label{chiA}
\tilde{\chi}_{\{\lambda\}}(S)=\frac{1}{2^l} \sum_{{\{\mu_i\}}} 
 \langle \hat{O}^{\{\!\mu_i\!\}}_{S} \rangle
   \hat{O}^{\{\!\mu_i\!\}}_S \;,
\end{equation}
and the sum in eq.~(\ref{chiA}) is restricted to the operators $\hat{O}^{\{\!\mu_i\!\}}_S$ that anti-commutes with $P_z$

Having the expression of the reduced density matrix obtained for a generic ground state projected in a generic finite subset $S$ we can turn back to 
our problem of the local distinguishability. 
From the basic concept of the distinguishability we can say that two state will be locally distinguishable if there exist a subset $S$ for which
the two reduced density matrices obtained projecting the two spin into $S$ are different. 
Hence a quantity that measure the distance between the two matrices, it is also a measure of their distinguishability.
There are several measures of distance between two matrices. 
In the present work we have decided to work with the trace distance~\cite{Nielsen2000} that allows to simplify our analysis.   
From the definition of the trace distance it is easy to recover that the maximum distance between two local density matrices of the form of 
eq.~(\ref{reduced3}) is reached when one state is one of the symmetric ground states and the other is one of the maximally symmetry broken ground 
states.
Named $\rho^{max}_{\{\lambda\}}(S)$ the reduced density matrix obtained projecting one of the maximally symmetry breaking ground states on $S$ and 
$D_S$ the maximum of the local distinguishability given by $D_S=\| \rho_{\{\lambda\}}^{max}(S) -\rho^{sym}_{\{\lambda\}}(S) \| $ it is easy to recover 
that $D_S$ is equal to 
\begin{equation}
\label{distance}
\!\!D_S\!=\!\frac{1}{2}\sum_{i=1}^{2^l} |\nu_i|\;,
\end{equation}
where the sets of the $\{\nu_i\}$ are the sets of the eigenvalues of the traceless matrix $\tilde{\chi}_{\{\lambda\}}(S)$

For a system in a magnetically ordered phase in static conditions, the fact that there exists a magnetic order parameter implies that $D_S$ is always 
greater than zero, regardless the choice of $S$. 
Viceversa for other kind of order, as the nematic phase~\cite{Lacroix2011,Giampaolo2015} in which the symmetry broken order parameter has a support 
greater than one single spin, the fact that  the $D_S$  is equal or different from zero will depend on the choice of $S$.

Till now we have considered the static problem in which there is no dependence on time. 
However our results can be easily generalized to the time dependent situations where the evolution is due to a sudden quench of the set
$\{\lambda\}$, from $\{\lambda_0\}$ to $\{\lambda_1\}$. 
In such case not only $H_{\{\lambda_0\}}$ and $H_{\{\lambda_1\}}$ will commute with the parity but also the operator 
$U_{\{\lambda_1\}} =\exp (-\imath H_{\{\lambda_1\}} t)$ will do the same. 
This implies that the time evolution induced by $U_{\{\lambda_1\}}$ does not change the superposition coefficient in eq.~(\ref{groundstate}).
Therefore, to evaluate the time dependent distance between the two matrices, it is enough to determine the behavior of the sum of the 
absolute value of the eigenvalues of $\tilde{\chi}_{\{\lambda_1\}}(S,t)$ that, in turn, is a function of the time dependent anti-symmetric spin 
correlation functions with support in $S$. 
Hence, also when the system is evolving, as a consequence of a sudden quench of the Hamiltonian parameters, the two ground state for 
which the distance reaches the maximum are, as in the stationary case, one of the maximally symmetry broken and one of the symmetric 
ground states.
We can hence generalize the results of the static case to the dynamic one and define, for any finite subset $S$, the time dependent maximum 
distance $D_S(t)$ as
\begin{equation}
\label{distance_time}
\!\!D_S(t) = \!\frac{1}{2}\sum_{i=1}^{2^l} |\nu_i(t)|
\end{equation}
where $\nu_i(t)$ are the time dependent eigenvalues of $\tilde{\chi}_{\{\lambda_0,\lambda_1\}}(S,t)$ defined as
\begin{equation}
\label{distance_time2}
 \tilde{\chi}_{\{\lambda_0,\lambda_1\}}(S,t)=\rho^{max}_{\{\lambda_0,\lambda_1\}}(S,t) \nonumber
 -\rho^{sym}_{\{\lambda_0,\lambda_1\}}(S,t) \;.
\end{equation}
In eq.~(\ref{distance_time2}) $\rho^{\max}_{\{\lambda_0,\lambda_1\}}(S,t)$ and $\rho^{sym}_{\{\lambda_0,\lambda_1\}}(S,t)$ are the straightforward 
generalization to the dynamic case of the reduced density matrix $\rho^{max}_{\{\lambda\}}(S)$ and $\rho^{sym}_{\{\lambda\}}(S)$.


\section{Numerical results for the $XY$ model}

 \begin{figure}[t]
    \includegraphics[width=8.5cm]{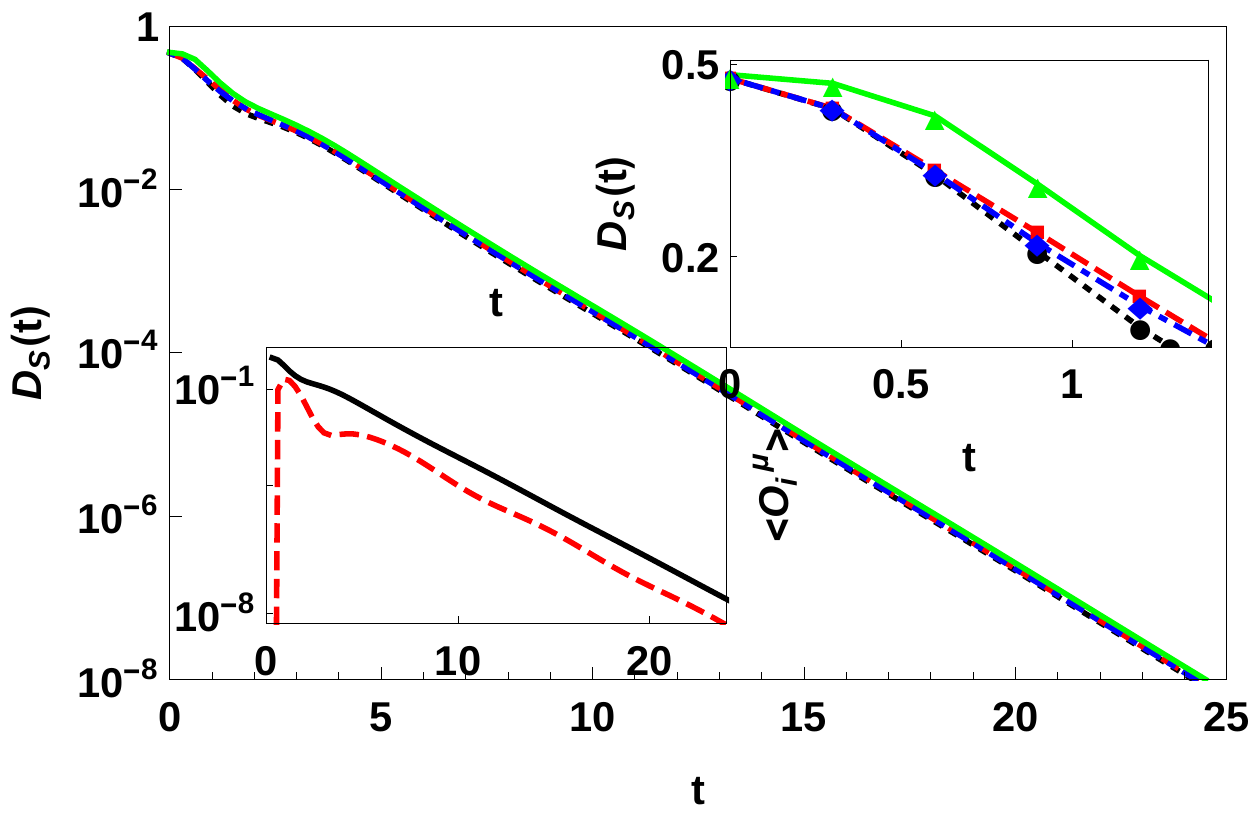}
    \includegraphics[width=8.5cm]{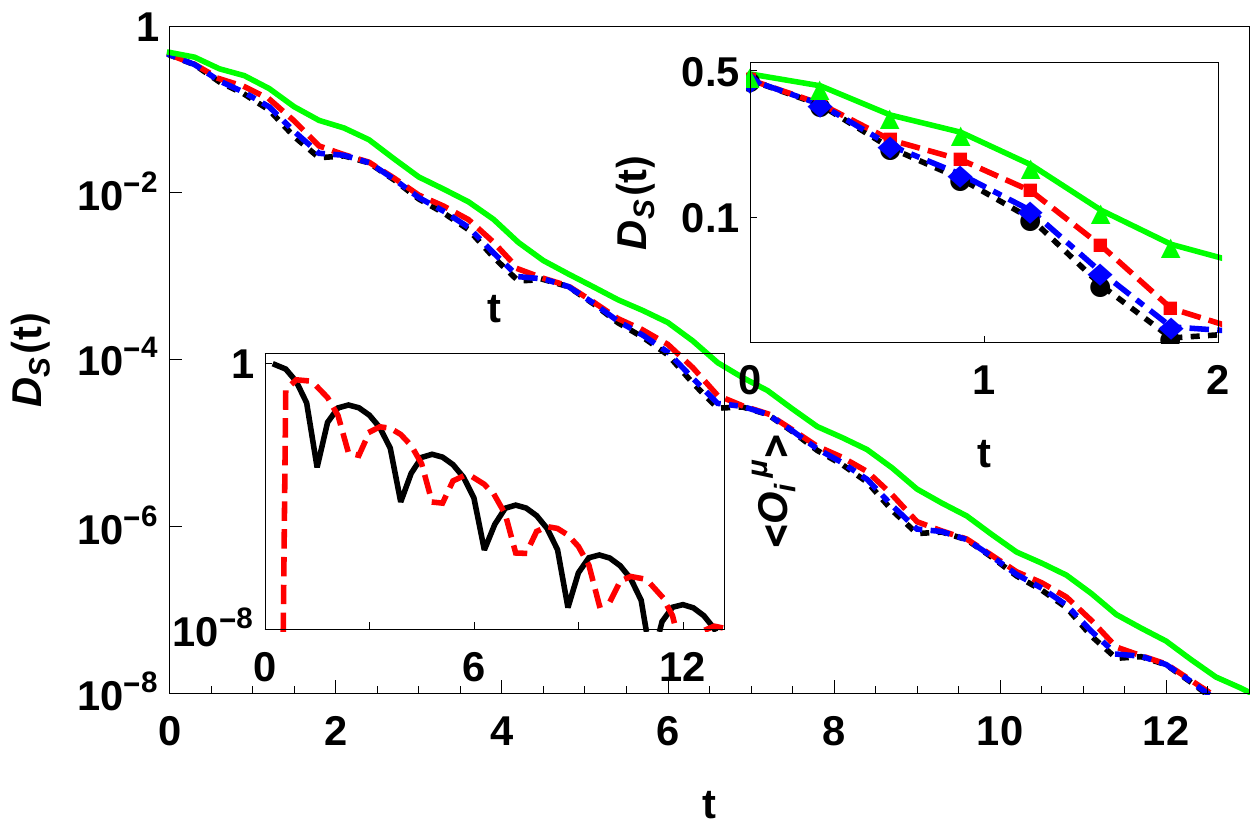}
     \caption{(Color online) Behavior of the distinguishability for two different sudden quenches for the $XY$ model. 
     In the upper panel we report our result for the case in which with $\gamma_0=\gamma_1=0.5$ the external field is quenched from 
     $h_0=0.2$ to $h_1=0.8$ while in the lower panel, always taking $\gamma_0=\gamma_1=0.5$ we quench the external field from 
     $h_0=0.4$ to $h_1=1.2$.
     In both case in the main plot we can see the behavior of the maximal distance $D_S(t)$ as function of the time $t$ 
     for $S$ made by one single spin (Black dotted line), 
     two neighbors spins (Blue dot-dashed line), two next neighbor spins (Red dashed line) and three spins (Green lines).
     In the inset at the top right we can see a zoom of the main inset for very short times in which the transient is highlighted.
     In the inset at the bottom left we show the behaviors, as function of time $t$, of the absolute value of the magnetization along the 
     $x$ $\langle \sigma_i^x \rangle$ (Black line) and along $y$ $\langle \sigma_i^y \rangle$ (Red line)
      }
    \label{figure1}
  \end{figure}
In the previous section we have defined a quantitative approach to the the local distinguishability.
Thanks to our approach we have proved that, for any subset $S$ and any time $t$, the maximal distance between the reduced density matrix obtained 
projecting into $S$ the image of two different ground states after a sudden quench is given in eq.~(\ref{distance_time}).
Hence we can start to study the time dependent local distinguishability for different one dimensional models.
The first model on which we focus our attention is the well known one-dimensional spin-1/2 $XY$-model with ferromagnetic nearest-neighbor 
interactions in the presence of a transverse magnetic field. 
Such model is described by the following Hamiltonian
\begin{equation}
  \label{XY_start}
\!\! \!\! H_{\{\gamma,h\}}\!\!=\!\!-\!\!\sum_j\!\!\left(\frac{1\!+\!\gamma}{2}\sigma^{x}_j\sigma^{x}_{j+1}\!+\!\frac{1\!-\!\gamma}{2}\sigma_j^{y}
\sigma^{y}_{j+1}
\right)\!\! -h\!\sum_j\!\sigma_j^{z} ,
 \end{equation}
where $\sigma_i^\zeta$ $(\zeta= x, \, y,\, z)$ are the standard Pauli spin-1/2 operators defined on the $i$-th spin of the chain and
the two parameters are respectively the anisotropy $\gamma$ and the transverse external field $h$.
Regardless the value of these two parameters, the Hamiltonian in eq.~(\ref{XY_start}) always commutes with the parity operator along $z$.
However, as it is well known, such model shows a magnetically ordered phase for $\gamma \in (0, 1]$ and 
$h<h_c\equiv 1$~\cite{Sachdev2000,Barouch1970} in which the parity symmetry is broken. 
In the magnetically ordered phase the ferromagnetic order parameter is given by
\begin{equation}
\label{xyorder}
\braket{\sigma_i^x} =\frac{2[\gamma^2 (1-h^2)]^{1/8}}{[2(1+\gamma)]^{1/2}} \; .
\end{equation}
We are, therefore, in the hypothesis that we have considered in Sec.II and hence we may apply the results presented there to evaluate the 
maximal quantum distinguishability.

\begin{figure}[t]
    \hspace{8.5cm}
    \includegraphics[width=8cm]{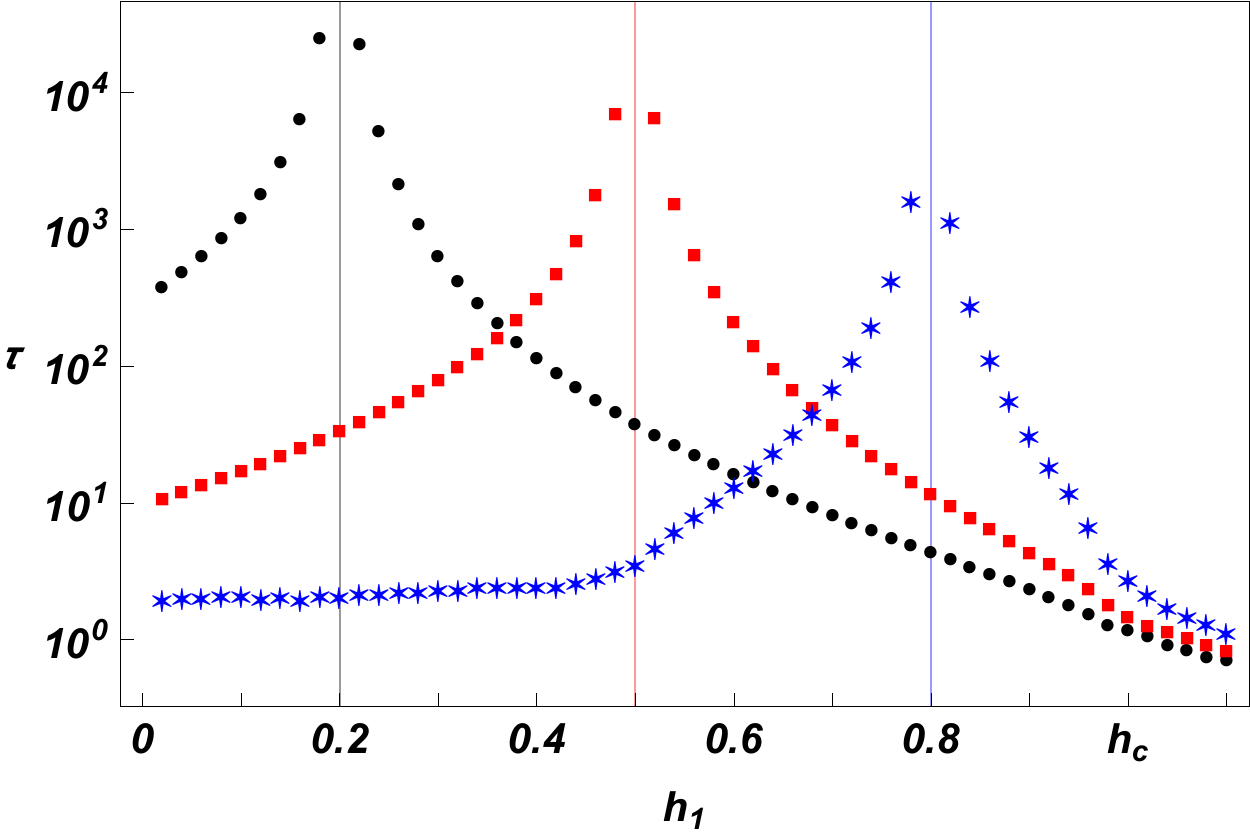}
     \caption{Time scale $\tau$, as function of the final external field $h_1$, of the exponential decay of $D_S(t)$ for a quench in the $XY$ model 
     that involves only the external field for several sets of initial parameters. 
     Black circles represent the case in which $\gamma_0=\gamma_1=0.8$ and the initial value of the external 
     field is $h_0=0.2$; red squares stand for the case in which $\gamma_0=\gamma_1=0.5$ and $h_0=0.5$; 
     blue stars represent the case in which $\gamma_0=\gamma_1=0.2$ and $h_0=0.8$.
    }
    \label{figure2}
  \end{figure}

As we have just said in Sec. II as a consequence that the order parameter never vanishes in the magnetically ordered phase also the maximal 
distinguishability in the static condition $D_S$ is always different from zero regardless the choice of $S$.
What happens after the quench?
We apply the methods described in the appendix to obtain the behavior of $D_S(t)$ for several choice of initial and final Hamiltonian parameters. 
The results, for two typical cases, are shown in Fig.~(\ref{figure1}).  
Even if we may have a short transient in which the maximum local distinguishability can increase respect to the one of the static case state, after a 
while $D_S(t)$ start to decay exponentially in time.
The presence, duration and relevance of the transient depends on the difference between the initial and final sets of the 
Hamiltonian parameters. 
Increasing the distance the transient becomes harder and harder to be seen. 
Moreover it is always more and more relevant as the size of $S$ increase. 
However, after the transient, all the maximal distinguishability $D_S(t)$ show an exponential decay $e^{-t/\tau}$ which a common time scale $\tau$.
The time scale does not depend on $S$ but depends on the parameters of the system before and after the quench and increase as the set of the 
parameter becomes closer and closer. 
As en example in Fig.~(\ref{figure2}) we have reported the behavior of the time scale, in the case of a quench that involves only the external field, 
as function of $h_1$ for several possible choice of $\gamma_0=\gamma_1$ and $h_0$. 
The exponential decay of $D_S(t)$ is a consequence of the exponential decay, with the same time scale, that characterize all the correlation 
functions with support included in $S$ that does not commute with the parity. 
The exponential behavior of some of these correlation functions are depicted in the bottom right insets of Fig.~(\ref{figure1}).
As it is possible to see by the insets, the time evolution induced by the quench of the Hamiltonian parameters realizes a magnetization along $y$ 
that in the static condition is equal to zero and that vanishes in the limit of large times.
A similar behavior is shared by all the parity breaking correlation functions that in the static situation vanish. 

As a consequence in the limit of diverging time, i.e. when $t\rightarrow \infty$, all the correlation functions which operator does not commute with 
the parity becomes zero.
This implies that, regardless the specific choice of a finite subset $S$, in the steady state, i.e. for very large time,  
the system loses completely any information about the superposition properties of the initial ground state. 
For any $S$ the reduced density matrices will commute with $P_z$.
Extending the analysis to the correlation functions which operator commutes with $P_z$, we note that if they are zero for $t\le t_0$, as for 
example $\langle \sigma_i^x\sigma_{i+1}^y\rangle$, they become different from zero soon after the quench but turn to be zero in the steady state 
with a behavior slower than the exponential one. 
This fact implies that, for any subset $S$, the reduced density matrix of the steady state $\rho_{\{\gamma,h\}}(u,v,S,t\rightarrow \infty)$ holds the 
same symmetries of the reduced density matrix obtained from the symmetric ground state in the stationary condition $\rho^{sym}_{\{\gamma,h\}}(S)$. 

However, the two states show very different physical properties due to the disappearance of the long range order implied by the absence of a non
vanishing order parameter. 
The most relevant example of such differences is the value of the mutual information between two very distant spins. 
In fact it is known~\cite{Hamma2016} that the symmetric ground states in a ferromagnetic phase are characterized by a non vanishing mutual information 
between two very far spin that is associated to the presence of non zero order parameter. 
But, as we have seen, after the quench, all the correlation functions that break the parity symmetry, hence including also the order parameter, go 
rapidly to zero, so implying the disappearance of the mutual information. 
This represent a further proof of the fragility of the states with global entanglement, detected by the persistence of a non vanishing
mutual information in the limit of large distance between the spins~\cite{Hamma2016}.

\section{Numerical results for the $N$-cluster Ising model}

In the previous section we have analyzed the time dependent local distinguishability for $XY$ models subjected to a sudden quench. 
We have seen that, regardless the particular choice of the initial and final set of parameters, as well as of the subset $S$, the maximally 
distinguishability goes to zero exponentially in time. 
As a consequence in the limit of very large time the system loses completely any information about the particular ground state it was before the 
quench. 
At this point the question that naturally comes in mind is: how general is this picture?

In the attempt to provide an answer to such question we decide to extend our analysis to a different spin-$1/2$ one dimensional model with a 
magnetically ordered phase.
Among others, we decide to focus on the family of models known as $N$-cluster Ising models~\cite{Smacchia2011,Giampaolo2014,Giampaolo2015}.
The reasons of this choice has to be found in the fact that: a) such family of models can be solved with the same approach used for the $XY$ model 
and showed in the appendix; b) The family of models presents a larger class of symmetries ($\mathbb{Z}_2^{\otimes N+1}$ instead of $\mathbb{Z}_2$)
Especially the second point makes these models really interesting to be analyzed. 
In fact in the magnetic phase only the $P_z$ symmetry results to be violates.
Therefore, differently from the $XY$ models, in the symmetry broken ground states not all the symmetries of the Hamiltonian are violated by the 
symmetry broken ground states.
This fact, as we will see soon, plays an important role in the dynamic of the $N$-cluster Ising models after a quench of the Hamiltonian parameters.
  
To start the analysis of such models let us introduce its Hamiltonian that reads
\begin{equation}
  \label{cluster_start}
\!\! H_{\{\varphi\}}^N\!\!=\!\!-\cos\varphi \sum_j\!\!{\sigma^{x}_j}Z_i^N{\sigma^{x}_{j+N+1}} + \sin \varphi \sum_j\!\! {\sigma_j}^{y}
{\sigma^{y}_{j+1}}.
 \end{equation}
Here $\varphi$ is the parameter that control the relative weight of the cluster term (the first sum of the r.h.s.) and of the Ising one while
the operator $Z_i^N$ stands for
\begin{equation}
  \label{cluster_operator}
 Z_i^N=\bigotimes_{k=1}^N \sigma^{z}_{i+k}\;.
 \end{equation}
It is easy to verify that the Hamiltonian in eq.~(\ref{cluster_start}) always commutes with the parity operator along $z$.
However it is well known that, regardless the size of the cluster interaction, such model shows an anti-ferromagnetic phase 
for $\varphi \ge \varphi_c\equiv \pi/4$ ~\cite{Giampaolo2015} in which the order parameter is equal to
\begin{equation}
\label{clusterorder}
(-1)^i  \braket{\sigma_i^y} =\left(1-\tan(\varphi)^{-2}\right)^{\frac{N+2}{8}} \; .
\end{equation}
Hence we are in the range of validity of the results for the distinguishability shown in Sec.II. 
Therefore we may extend to such model the same analysis made made for the $XY$ model in the previous section.
 \begin{figure}[t]
    \includegraphics[width=8.5cm]{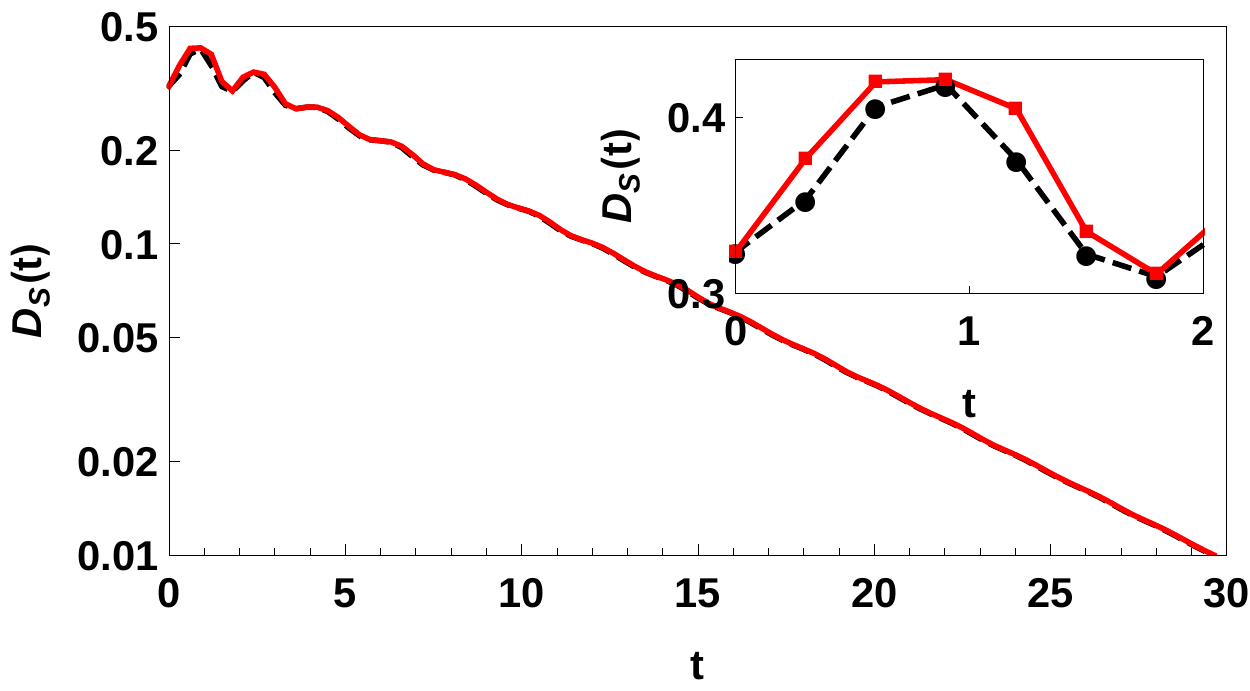}
    \includegraphics[width=8.5cm]{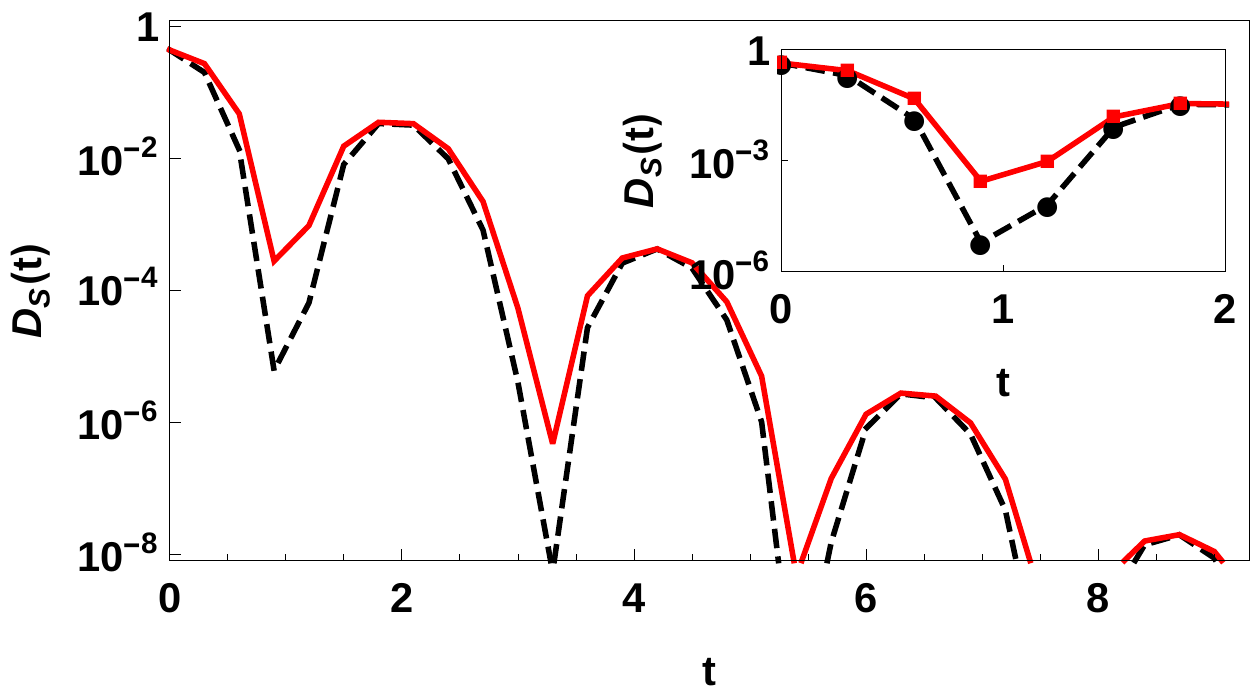}
     \caption{(Color online) Behavior of the distinguishability for two different sudden quenches for the $N$-cluster Ising model with $N=1$. 
     In the upper panel we report our result for the case in which $\varphi$ is sudden quenched from $\varphi_0=\frac{5}{16}\pi$ to 
     $\varphi_1=\frac{7}{16}\pi$ while
     in the lower panel, is quenched from $\varphi_0=\frac{3}{8}\pi$ to $\varphi_1=\frac{1}{8}\pi$ .
     In both case in the main plot we can see the behavior of the maximal distance $D_S(t)$ as function of the time $t$ 
     for $S$ made by one single, two neighbors spins and two next neighbor spins (Black dotted line), and three spins (Red solid line).
     In the inset we can see a zoom of the main inset for very short times in which the transient is highlighted.
      }
    \label{figure3}
  \end{figure}
  
In Fig.~(\ref{figure3}) we show the numerical results for the time dependent maximally distinguishability $D_S(t)$ for two different quenches of the 
parameter $\varphi$.
Comparing the results in  Fig.~(\ref{figure3}) for the $N$-cluster Ising with the ones for the $XY$ model in Fig.~(\ref{figure1}) we may see several 
analogies and differences.

As for the $XY$ model the $D_S(t)$ for the $N$-cluster Ising presents a transient in which the distinguishability may increase. 
Furthermore, in analogy with the results shown in Fig.~(\ref{figure1}), after the transient the maximally distinguishability show an exponential 
decay, independently of the value of $\varphi$ before and after the quench and the particular choice of $S$.
Also in this case the time scale does not depend on $S$ but depends on the parameters of the system before and after the quench and increase 
as the difference between $\varphi_0$ and $\varphi_1$ decreases. 
Therefore also in this second family of models, in the steady state realized at very large time, all the informations about the particular ground 
state are completely lost.

But, nevertheless these analogies, the presence of the other symmetries that are not violated even in the symmetry broken states plays
an extremely important role. 
To explain this role we have to enter in some details of the evaluation process explained in the appendix.
To evaluate all the correlation functions we use the following approach. 
At the very beginning, using the Jordan-Wigner~\cite{Jordan1928} transformation we turn the spin operator associate to the correlation function to a 
fermionic one.
Hence we use the Wick Theorem~\cite{Wick1950} to obtain the expression of the spin correlation functions in term of two body fermionic correlation 
functions named $f(r,t)$, $g(r,t)$ and $h(r,t)$ where $r$ is an index related to the distance between the two operators.
In the static case for the $N$-cluster Ising model, as well as for the $XY$ model, we have that $f(r,0)=-h(r,0)=\delta_{r,0}$ where 
$\delta_{r,0}$ is equal to one when $r=0$ and zero otherwise. 
On the other hand the presence of the large group of symmetries that characterized the $N$-cluster models implies that
$g(r,0)=0\; \forall r\neq a (N+2)+1$ where $a$ is an integer~\cite{Smacchia2011,Giampaolo2015}. This behavior for the $g(r,0)$ is completely 
different with what happens in the $XY$ case where all the $g(r,0)\neq 0$.
Such a difference is related to the fact that the $N$-cluster Ising model holds a large class of symmetries that implies that, for example, 
the magnetization along $z$ is always equal to zero for all ground states of the Hamiltonian in eq.~(\ref{cluster_start}).
Not only. 
Even more important for our analysis, is that the only spin correlation functions with support included in a subsystem $A$ with a size $l<N+2$
that is different from zero in the static condition can be $\langle\sigma_i^y\rangle$. Consequently we have that 
$D_{S_1}=D_{S_2}$ if $l_{S_1},l_{S_2}<N+2$.

When the quench is take into account also the two body fermionic correlation functions start to depend on time. 
But, independently on the parameters of the quench, we still have at any time $t>0$ that $g(r,t)=0\; \forall r\neq a (N+2)+1$ while for 
$f(r,t)$ and $h(r,t)$ we have that $f(r,t)=-h(r,t)$ and $f(r,t)=0\; \forall r\neq a (N+2)$. 
As a consequence all the spin correlation functions which operators have a support in a subsystem with a size lower than $N+2$, anti-commute with 
$P_z$ and that were zero in the stationary condition, remain zero also after the quench. 
This is exactly the opposite of which happens in the $XY$ model in which all the spin correlation functions becomes different from zero soon after 
the quench, as can be seen looking at the bottom left insets of Fig.(\ref{figure1}).
As for the static case, also this results is due to the presence of the residual symmetry of the system that are not violated by the ground states 
that break the parity symmetry. 
As a consequence we have that we can generalize the previous result obtained in stationary case also in the dynamic one writing 
$D_{S_1}(t)=D_{S_2}(t)$ if $l_{S_1},l_{S_2}<N+2$.

\section{Conclusion}

In conclusion we have analyzed the problem of the local distinguishability between the different ground states of a magnetic ordered phase
after a quench of the Hamiltonian parameters. 
To have an useful tool for our goal, at the beginning, we have developed a generic quantitative approach to the problem of the distinguishability, 
and soon after we have applied our results to two different families of models that show different class of symmetries. 
The two families of models considered are the $XY$ models and the $N$-Cluster Ising models. 
Independently of the particular model, the finite subset taken into account and the parameter before and after the quench, we have that, after a 
short transient,the distinguishability becomes to disappear exponentially in time. 
The informations on the superposition parameters are completely erased by the time evolution, even though all the models we have examined are 
integrable models and therefore local stationary states that are realized after a very long time kept informations on the initial and final 
parameters of the system. 
In other words the information about the superposition is lost even if the system does not thermalize.

Moreover we provide the proof that if the symmetry broken by the magnetic order is the the unique local symmetry present in the Hamiltonian, as in the 
$XY$ model, an unitary time evolution induced by the quench force the rise of long range correlation functions also in the direction of minimum 
asymmetry. 
These long range correlation functions may induce interesting phenomena such as the amplification of the entanglement between two neighbors spins 
that may have relevant applications for the quantum information and computation~\cite{Bayat2013}. 

Our findings also provide further evidence of the fragility of the states that show a nonzero global entanglement. 
It is in fact well known that these states are unstable from a point of view of interactions with an external environment, as it is shown, for example, 
by the behavior of the local convertibility~\cite{Cianciaruso2014} or of the mutual information~\cite{Hamma2016}. 
But we show that they are unstable even in the presence of a unitary evolution, typical of a closed system and not interacting with the outside. 
It is however important to remember that our results were obtained in the context of a short-range one-dimensional model which, as is well 
known~\cite{Mermin1966}, does not allow phase transitions at temperatures different from zero. 
In a future work we will try to understand how our results can be generalized at models that show ordered phase even at temperatures different 
from zero.

Our results are not the first about the time-evolution of symmetry broken ground states. 
We have, however, provided a more general approach based on all the correlation functions that break the symmetry and not only based on the analysis 
of the time evolution of the order parameter.
In this sense our work can be seen as a generalization of some previous results, obtained in the framework of the $XY$ model by the group 
leaded by P. Calabrese~\cite{Calabreseworks}.

\section*{acknowledgments}  
We wish to thank A. Hamma and V. E. Korepin for the interesting discussions and for reading an early version of the letter. 
S.M.G. acknowledges financial support from the Ministry of Science, Technology and Innovation of Brazil while G. Z. thanks 
EQuaM - Emulators of Quantum Frustrated Magnetism, Grant Agreement No. 323714


\appendix

\section{Analytic Approach to the problem of the quench}

In this appendix we illustrate in details the method that we have used all along the paper to evaluate the time dependent correlation functions with 
which we may reconstruct the reduced density matrices and hence evaluate $D_S(t)$.
Such approach can be used for all the models that can be solved using Jordan-Wigner transformations and can be generalized to all the possible time 
dependences that preserve the parity symmetry of the Hamiltonian.

As we have seen in the Sec.II of the paper, all the correlation functions that we used in the evaluation can be obtained from expectation values 
of properly chosen operators on a symmetric ground state. 
This is a very important point because using Jordan-Wigner transformation, the symmetric ground states are the only ones that can be easily obtained. 
Our approach can be considered divided in three parts.
In the first we evaluate the symmetric ground states at rest, i.e. before the quench. 
In the second we apply the time evolution and we obtain the image of the symmetric ground state as function of the time.
In the third we extract all the time dependent correlation functions that we need.

\subsection{The static ground state}

In our work we focus on two different families of models: the $XY$ model and the $N$-cluster Ising models.
Both the two families of models can be analytically diagonalized using the Jordan-Wigner transformations
\begin{equation}
\label{JWT}
 c_j  = \prod_{k=1}^{j-1} \left( \sigma^z_k \right) \sigma_j^- 
\; ; \; \;\; \; c_j^\dagger  = \prod_{k=1}^{j-1} \left( \sigma^z_k \right) \sigma_j^+ \;,
\end{equation}
that map spin-$1/2$ systems into a noninteracting fermions moving freely along the chain only obeying Pauli's exclusion principle.
Here $c_j$ and $c_j^\dagger$ stand respectively for the annihilation and creation fermionic operators in the $j$-th site
The fermionic problem can be diagonalized using a Fourier transformation
\begin{equation}
\label{FT1}
 b_k = \frac{1}{\sqrt{N}} \sum_{j} c_j e^{-ikj} \; ; \; \;\; \;
 b_k^\dagger  = \frac{1}{\sqrt{N}} \sum_{j} c_j^\dagger e^{ikj} \;,
\end{equation}
where $k = 2\pi l/N$ and $l$ is an integer index that runs from -$N/2$ to $N/2$ where $N$ is the total number of spins in the chain. 

In all the models studied in the paper we obtain that the Hamiltonian can be expressed as 
\begin{equation}
\label{blockHamiltonian}
 H_{\{\lambda\}}=\sum_{k>0} \tilde{H}_{\{\lambda\},k} \;,
\end{equation} 
where $ \tilde{H}_{\{\lambda\},k}$ is a term acting only on fermions with momentum $k$ and $-k$. 
This local, in the momentum space, Hamiltonian is equal to 
\begin{eqnarray}
  \tilde{H}_{\{\lambda\},k} &= & 2\varepsilon_{\{\lambda\},k} (b_{k}^\dagger b_{k}+b_{-k}^\dagger b_{-k}-1) \nonumber \\
   & + &  2\imath\delta_{\{\lambda\},k}( b_{k}^\dagger b_{-k}^\dagger - b_{-k} b_{k} )\; ,
\end{eqnarray}
where $\varepsilon_{\{\lambda\},k}$ and $\delta_{\{\lambda\},k}$ will depends on the model under analysis. 
In our case we have  
\begin{eqnarray}
 \label{epsdeltaxy}
 \varepsilon_{\{\lambda\},k}  &=&\cos(k)-h \nonumber \\ 
 \delta_{\{\lambda\},k}  &=&\gamma \sin(k)\;.
\end{eqnarray}
for the $XY$ model and to 
\begin{eqnarray}
 \label{epsdeltacluster}
 \varepsilon_{\{\lambda\},k}  &=&\cos((N+1)k)\cos{\varphi}-\cos(k)\sin(\varphi) \nonumber \\ 
 \delta_{\{\lambda\},k}  &=&\sin((N+1)k)\cos{\varphi}-\sin(k)\sin(\varphi) \;.
\end{eqnarray}
for the $N$-cluster Ising models.

%

Hence, thanks to the Jordan-Wigner and Fourier transformations, the spin Hamiltonians are mapped in sums of non-interacting four level systems 
$\tilde{H}_{\{\lambda\},k}$, each one of them acting only on fermionic states with wave number equal to $k$ or -$k$. 
Defining the occupation number basis $\ket{1_k,1_{-k}}$, $\ket{0_k,0_{-k}}$, $\ket{1_k,0_{-k}}$, $\ket{0_k,1_{-k}}$, each $\tilde{H}_{\{\lambda\},k}$ 
corresponds to a $4\times4$ matrix 
\begin{equation}
\label{Hamiltonian_matrix}
\tilde{H}_{\{\lambda\},k}=\left( 
\begin{array}{cccc}
2 \varepsilon_{\{\lambda\},k} & 2 \imath \delta_{\{\lambda\},k} & 0 & 0 \\
-2 \imath \delta_{\{\lambda\},k} & -2 \varepsilon_{\{\lambda\},k} & 0 & 0 \\
 0 & 0 & 0 & 0 \\
 0 & 0 & 0 & 0 
\end{array}  
\right)  \; .
\end{equation}
From this expression of $\tilde{H}_{\{\lambda\},k}$ it is easy to evaluate the ground state energy, that results to be
\begin{equation}
\label{XY_Ek0}
 \omega_{\{\lambda\},k}=- 2 \sqrt{\varepsilon_{\{\lambda\},k}^2+\delta_{\{\lambda\},k}^2};.
\end{equation}
The associated ground state $\ket{\psi_{\{\lambda\},k}}$ is a superposition of $\ket{1_k,1_{-k}}$ and $\ket{0_k,0_{-k}}$
\begin{equation}
\label{ground_state_k_static}
 \ket{\psi_{\{\lambda\},k}}=\alpha_{\{\lambda\},k} \ket{1_k,1_{-k}} + \beta_{\{\lambda\},k} \ket{0_k,0_{-k}}\;.
\end{equation}
where the superposition parameters are given by
\begin{eqnarray}
 \label{XY_alfa_beta}
\!\!\!\!\!\!\!\!
\alpha_{\{\lambda\},k} \!&\!=\!&\! \imath \frac{\varepsilon_{\{\lambda\},k} \!-\!\sqrt{\varepsilon_{\{\lambda\},k}^2+\delta_{\{\lambda\},k}^2}}
{\sqrt{\delta_{\{\lambda\},k}^2 +\left(\varepsilon_{\{\lambda\},k}\! -\!\sqrt{\varepsilon_{\{\lambda\},k}^2+\delta_{\{\lambda\},k}^2} \right)^2}} 
  \\
\!\!\!\!\!\!\!\! \beta_{\{\lambda\},k}\! &\!=\!&\!  \frac{\delta_{\{\lambda\},k}}{\sqrt{\delta_{\{\lambda\},k}^2
 +\left(\varepsilon_{\{\lambda\},k} \!-\!\sqrt{\varepsilon_{\{\lambda\},k}^2+\delta_{\{\lambda\},k}^2} \right)^2}}\nonumber \;.
\end{eqnarray}

Since the Hamiltonian is the sum of the non-interacting terms $\tilde{H}_{\{\lambda\},k}$, each one of them acting on a different Hilbert space, 
the ground state of the total Hamiltonian will be a tensor product of all $\ket{\psi_{\{\lambda\},k}}$ 
\begin{equation}
 \label{globalgs_static}
 \ket{\psi_{\{\lambda\}}}=\bigotimes_k\ket{\psi_{\{\lambda\},k}} \;.
\end{equation}
It is worth to note that the ground state so defined holds a well defined symmetry that depends on the particular set of parameters taken into 
account~\cite{Blasone2010}. 
However, going towards the thermodynamic limit the energy gap between the even and the odd sectors tends to vanish and when the number of spins in the 
system diverges we have a perfect degeneracy, below the quantum critical point, between an even and an odd ground 
state~\cite{Barouch1970,Sachdev2000}.

\subsection{Time evolution induced by a sudden quench of the Hamiltonian parameters}

At this point we have obtained an analytical expression of the symmetric ground state before the quench.
Let us now move to describe the dynamics of a ground state after a sudden change of the set of the Hamiltonian parameters between $\{\lambda_0\}$ and
$\{\lambda_1\}$.
For any time $t \ge 0$ the system will be described by the state $\ket{\psi(t)}=U(\{\lambda_1\},t) \ket{\psi_{\{\lambda_0\},k}}$ where 
$U(\{\lambda_1\},t)=e^{-i H_{\{\lambda_1\}} t}$ is the time evolution unitary operator. 
However, taken into account that: 
1) the wave number $k$ does not depends on the set of the Hamiltonian parameters $\{\lambda\}$ and hence eq.~(\ref{blockHamiltonian}) is still valid; 
2) the initial state $\ket{\psi_{\{\lambda_0\}}}$ can be written as tensor product of states defined on each single $k>0$ 
(eq.~(\ref{globalgs_static})), 
we obtain
\begin{equation}
 \label{globalgs_dynamic}
 \!\!\!\ket{\psi(t)}=\bigotimes_k\ket{\psi_k(t)}=\bigotimes_k U_k({\{\lambda_1\}},t)\ket{\psi_{{\{\lambda_0,\}},k}} \;.
\end{equation}
$U_k({\{\lambda_1\}},t)$ is a time evolution operator that acts on the subset made by the momenta $k$ and $-k$.
The explicit expression of operator $U_k({\{\lambda_1\}},t)$ can be determined  by the Heisenberg equation 
\begin{equation}
\label{Heisenberg_equation}
 \imath \frac{d}{dt}U_k({\{\lambda_1\}},t) = \tilde{H}_{\{\lambda_1\},k} U_k({\{\lambda_1\}},t) \;.
\end{equation}
As $\tilde{H}_{\{\lambda_1\},k}$ also $U_k({\{\lambda_1\}},t)$ can be represented as a matrix which coefficients can be determined 
by the solution of eq.~(\ref{Heisenberg_equation}). 
From  eq.~(\ref{Heisenberg_equation}), taking into account eq.~(\ref{Hamiltonian_matrix}) we obtain two non trivial systems of coupled differential 
equations with constant coefficients. 
The first is given by 
\begin{equation}
\label{eqdif3}
\Bigg \{
\begin{array}{l}
\imath \dot U_{11,k}(t) = 2 \varepsilon_{\{\lambda_1\},k} U_{11,k}(t) - 2 \imath \delta_{\{\lambda_1\},k} U_{12,k}(t)\\
\imath \dot U_{12,k}(t) = 2 \imath \delta_{\{\lambda_1\},k} U_{11,k}(t) - 2 \varepsilon_{\{\lambda_1\},k} U_{12,k}(t) \nonumber
\end{array}\;,
\end{equation}
while the second is 
\begin{equation}
\label{eqdif4}
\Bigg \{
\begin{array}{l}
\imath \dot U_{21,k}(t) = 2 \varepsilon_{\{\lambda_1\},k} U_{21,k}(t) - 2 \imath \delta_{\{\lambda_1\},k} U_{22,k}(t)\\
\imath \dot U_{22,k}(t) = 2 \imath \delta_{\{\lambda_1\},k} U_{21,k}(t) - 2 \varepsilon_{\{\lambda_1\},k} U_{22,k}(t) \nonumber
\end{array}\; .
\end{equation} 
We can decouple the two systems of differential equations, and transform them into four second order differential equations, with constant 
coefficients that can be solved taking into account the opportune boundary conditions. 
We obtain for $U_{11,k}(t)$
\begin{equation}
\label{eqdifU11}
\left\{
\begin{array}{l}
\ddot U_{11,k}(t) +\omega_{\{\lambda_1\},k}^2 U_{11,k}(t)=0\\
U_{11,k}(0)=1\\
\dot U_{11,k}(0)=-2 \imath  \varepsilon_{\{\lambda_1\},k}
\end{array}
\right.
\end{equation}
for $U_{12,k}(t)$
\begin{equation}
\label{eqdifU12}
\left\{
\begin{array}{l}
\ddot U_{12,k}(t) +\omega_{\{\lambda_1\},k}^2 U_{12,k}(t)=0\\
U_{12,k}(0)=0\\
\dot U_{12,k}(0)=2 \imath  \delta_{\{\lambda_1\},k}
\end{array}
\right.
\end{equation}
for $U_{21,k}(t)$
\begin{equation}
\label{eqdifU21}
\left\{
\begin{array}{l}
\ddot U_{21,k}(t) +\omega_{\{\lambda_1\},k}^2 U_{21,k}(t)=0\\
U_{21,k}(0)=0\\
\dot U_{21,k}(0)=-2 \imath  \delta_{\{\lambda_1\},k}
\end{array}
\right.
\end{equation}
and, at the end, for $U_{22,k}(t)$
\begin{equation}
\label{eqdifU22}
\left\{
\begin{array}{l}
\ddot U_{22,k}(t) +\omega_{\{\lambda_1\},k}^2  U_{22,k}(t)=0\\
U_{22,k}(0)=1\\
\dot U_{22,k}(0)=2 \imath  \varepsilon_{\{\lambda_1\},k}
\end{array}
\right.
\end{equation}
Solving the above differential equations we have 
\begin{eqnarray}
 \label{EvOp}
\!\!\!\!U_{11,k}(t)\!&\!=\!&\!\cos(\omega_{\{\lambda_1\},k} t)
-\imath  \frac{\varepsilon_{\{\lambda_1\},k}}{\omega_{\{\lambda_1\},k}}\sin(\omega_{\{\lambda_1\},k}t) \nonumber\\
\!\!\!\!U_{12,k}(t)\!&\!=\!&\imath  \frac{\delta_{\{\lambda_1\},k}}{\omega_{\{\lambda_1\},k}}\sin(\omega_{\{\lambda_1\},k} t)\\
\!\!\!\!U_{21,k}(t)\!&\!=\!&-\imath \frac{\delta_{\{\lambda_1\},k}}{\omega_{\{\lambda_1\},k}}\sin(\omega_{\{\lambda_1\},k} t) \nonumber \\
\!\!\!\!U_{22,k}(t)\!&\!=\!&\!\cos(\omega_{\{\lambda_1\},k} t)+\imath \frac{\varepsilon_{\{\lambda_1\},k}}{\omega_{\{\lambda_1\},k}}
\sin(\omega_{\{\lambda_1\},k} t)\nonumber \; .
\end{eqnarray}

With the explicit expression for the time evolution unitary operator $U_k(\{\lambda_1\},t)$ we may obtain, for any wave number $k$ and time $t$,
the image, of the initial state $\ket{\psi_{\{\lambda_0\},k}}$
\begin{equation}
\label{ground_state_k_dynamic}
 \ket{\psi_k(t)}=\tilde{\alpha}_{k}(t) \ket{1_k,1_{-k}} + \tilde{\beta}_{k}(t) \ket{0_k,0_{-k}}
\end{equation}
where
\begin{eqnarray}
\label{XY_At_Bt}
\tilde{\alpha}_k(t) & = & \alpha_{\{\lambda_0\},k} \cos( \omega_{\{\lambda_1\},k} t) \nonumber \\
& & -\imath   \frac{\varepsilon_{\{\lambda_1\},k} \alpha_{\{\lambda_0\},k}
        - \delta_{\{\lambda_1\},k} \beta_{\{\lambda_0\},k}}{\omega_{\{\lambda_1\},k}} \sin(\omega_{\{\lambda_1\},k} t) \nonumber \\
\tilde{\beta}_k(t) & = & \beta_{\{\lambda_0\},k} \cos(\omega_{\{\lambda_1\},k} t)\! \\
& & 
- \imath  \frac{\delta_{\{\lambda_1\},k} \alpha_{\{\lambda_0\},k} - \varepsilon_{\{\lambda_1\},k}  \beta_{\{\lambda_0\},k}}{\omega_{\{\lambda_1\},k}} 
\sin( \omega_{\{\lambda_1\},k} t) \nonumber
\end{eqnarray}
As in the static case, also for any time after a quench, the state $\ket{\psi_k(t)}$ still preserve the parity.
Knowing $\ket{\psi_k(t)}$ for any $k$, we hold the perfect knowledge of $\ket{\psi(t)}$. 
Hence the problem is to extract from such knowledge of the state at a generic time $t$, all the information that we need. 
In the following sections we illustrate how to evaluate all the correlation functions that we have used in the main text.

\subsection{Fermionic correlation functions}

At this point we have the expression of the state after a quench as a function of time. 
And from such expression we have to extract the spin correlation functions that allow us to reconstruct the different reduced density matrix.
However, the state is not expressed neither in terms of the spins nor in term of fermionic operators in the real space, but in terms of fermionic 
variables in the momentum space. 
Therefore the third step of our approach can be considered to be divided in two. 
In the first part, starting from eq.~(\ref{XY_At_Bt}) and taking into account the Fourier Transform in eq.~(\ref{FT1}), we obtain the different 
two body correlation functions between fermionic operators in real space. 
Soon after we use the knowledge of such correlation functions, and the Wick's Theorem~\cite{Wick1950} to reconstruct all the spin correlations.
To simplify such process it is convenient to introduce the Majorana fermionic operators~\cite{Lieb1961,Barouch1970} indicated, respectively, with 
$A_j$ and $B_j$ 
\begin{equation}
\label{ABoperators}
 A_j = c_j + c_j^\dagger \quad , \quad B_j = c_j - c_j^\dagger \; ,
\end{equation}
where $j$ is an index that runs on all the spins of the system. 
In general after this process we obtain a fermionic operator made by a large number of fermionic terms that can be evaluated applying the 
Wick's theorem.
Having two family of fermionic operators, it is enough to evaluate five types of expectation values that possess all the ingredients to 
determine each spin correlation function that we need. 

From the expression of $\ket{\psi_k(t)}$ in eq.~(\ref{ground_state_k_dynamic}) we can immediately seen that both $\braket{A_j}_t$ and $\braket{B_j}_t$ 
vanish (from now on, in the appendix, we use $\braket{O}_t$ as a shortcut for $\bra{\psi(t)} O \ket{\psi (t)}$ ). 
In fact, adding or removing a single fermion from $\ket{\psi_k(t)}$ the state is driven in an orthogonal subspace that implies that 
$\braket{A_j}_t=\braket{B_j}_t=0$. 
As a consequence, we have that if a spin operator is mapped into a fermionic operator made by an odd number of components, its expected value on 
$\ket{\psi (t)}$, that is an eigenstate of the parity operator $P_z$, vanishes.

On the contrary the other three basic elements, i.e. $f(i,k,t)=\braket{A_i A_k}_t$, $g(i,k,t)=\braket{B_i A_k}_t$ and $h(i,k,t)=\braket{B_i B_k}_t$
can be non-zero and must be evaluated if we want to obtain the explicit value of a generic spin correlation function. 
Before to start,let us note that, because we are considering a model that is invariant under spatial translation, also these functions must hold the 
same property and hence they have not to depend on the particular choice of the spins $i$ and $k$ but only on their relative distance $r=i-k$. 
Therefore we have that $f(i,k,t)=f(r,t)$, $h(i,k,t)=h(r,t)$ and $g(i,k,t)=g(r,t)$. 
At this point we are ready to begin the derivation of the different functions. Let us start our analysis with the $g(r,t)$ function that is defined as
\begin{equation}
 g(r,t) = \braket{ B_r A_0 }= \bra{\psi(t)} (c_r - c_r^\dagger) (c^\dagger_0 + c_0)  \ket{\psi (t)}
\end{equation}
After a long but straightforward calculation, we obtain
\begin{eqnarray}
\label{gr}
 g(r,t) &=&  \frac{2}{N} \sum_{k>0} \big[(|\tilde{\beta}_k(t)|^2 -|\tilde{\alpha}_k (t)|^2) \cos(kr) \\
 &  & \; \; \;\; \; \;\; \; \;\; \;+ i (\tilde{\alpha}_k^*(t) \tilde{\beta}_k (t)- \tilde{\alpha}_k(t) \tilde{\beta}_k^*(t) ) \sin(kr)\big] \nonumber
\end{eqnarray}
Following the same approach we obtain for $f(r,t)$ and $h(r,t)$
\begin{equation}
\label{fhr1}
f(r,t) = \delta_{r,0}+ \frac{i}{N} \sum_{k>0} (\tilde{\alpha}_k^* \tilde{\beta}_k + \tilde{\alpha}_k \tilde{\beta}_k^* ) \sin(kr) 
\end{equation}
\begin{equation}
\label{ghr1}
h(r,t) = -\delta_{r,0}+ \frac{i}{N} \sum_{k>0} (\tilde{\alpha}_k^* \tilde{\beta}_k + \tilde{\alpha}_k \tilde{\beta}_k^* ) \sin(kr)
\end{equation}
where $\delta_{r,0}$ is the Kronecker delta that is different from zero only when $r=0$. 

At this point, to obtain the results in the thermodynamic limit it is enough to substitute the sum over all $k>0$ $\sum_k$ with the normalized
integral over all $k$ between $0$ and $\pi$, i.e. $\int_0^\pi dk$

Taking $t=0$ the functions $f(r,t)$, $g(r,t)$ and $h(r,t)$ coincide with the static ones. 
In all the models that we have analyzed in the paper, because $\alpha_{\{\lambda\},k}$ is an imaginary number while $\beta_{\{\lambda\},k}$ is real
the elements in the second sum in the definition of both $f(r,0)$ and $h(r,0)$ are all zero. 
Hence, taking into account the normalization condition $|\alpha_{\{\lambda\},k}|^2+|\beta_{\{\lambda\},k}|^2=1$, we obtain 
$f(r,0)=-h(r,0)=\delta_{r,0}$. 
For $g(r,0)$ the situation is completely different. 
In the $XY$ model all $g(r,0)$ are, in general, different from zero~\cite{Barouch1970}.
On the contrary for the $N$-Cluster Ising model we have that only if $r$ satisfy the relation $r=a(N+2)+1$, where $a$ is an integer, we may have
$g(r,0)\neq0$~\cite{Smacchia2011,Giampaolo2015}.

For $t>0$, solving numerically the integrals, the difference between the two family of models increases.  
In fact, while for the $XY$ models all the functions becomes different from zero, this is not true in the case of the $N$-cluster Ising model.
In these models if we have that $g(r,0)=0$, due to the fact that the parity symmetry is not the unique local symmetry of the Hamiltonian, 
also $g(r,t)=0$. 
Not only.
The effect of the symmetries becomes also evident for the $f(r,t)$ and $h(r,t)$ functions. 
In fact we have that for any $t>0$ only when $r=a(N+2)$ we may have $f(r,t)=-h(r,t)\neq0$.
However, in all the models, when $t$ diverges we obtain again the same structure of the static case.

\subsection{Spin correlation functions}

\begin{figure}[t]
\includegraphics[width=8.5cm]{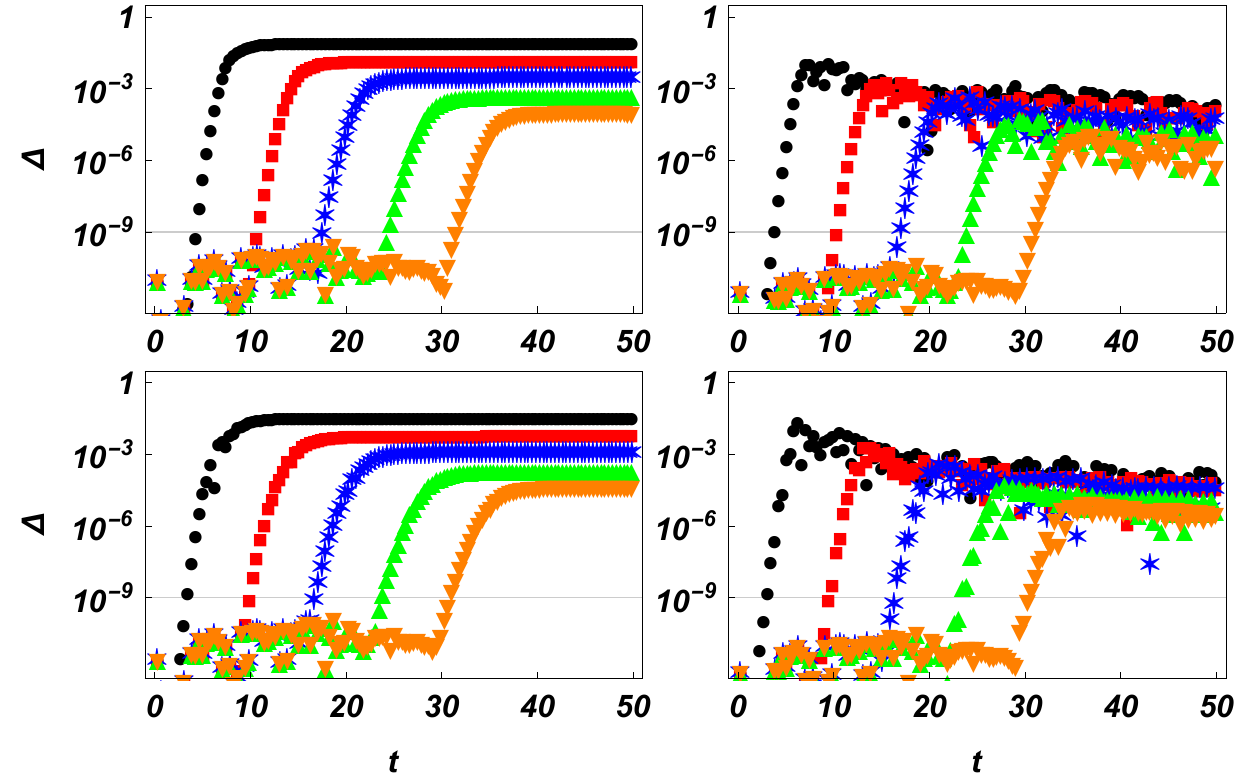}
\caption{(Color online) Behavior of the absolute values of the differences $\Delta$ between the values of the four broken-symmetry correlation 
functions obtained choosing $r_{max}=r+\Delta r$ and $r_{max}=r$, as function of time for a fixed set of the Hamiltonian parameters
$\gamma=0.8$, $h_0=0.2$ and $h_1=0.8$. We have arbitrarily chosen $\Delta r=10$. The different points stands for: black circles (left-most curve) 
$r=20$; red squares $r=40$; blue stars $r=60$; green upward triangles $r=80$; orange downward triangles (right-most curve) $r=100$; 
From the top left in the clockwise order we have plotted the difference for the following correlation functions $\braket{\sigma_i^x}$, 
$\braket{\sigma_i^y}$, $\braket{\sigma_i^y\sigma_{i+1}^z}$ and $\braket{\sigma_i^x\sigma_{i+1}^z}$. The lines at $10^{-9}$ indicate the border between
computational noise and significant differences}
\label{nsym_corr_diff}
\end{figure} 

With the knowledge of $g(r,t)$, $f(r,t)$ and $h(r,t)$ we may evaluate all the spin correlation functions at any time. 
However the approach to the evaluation is different depending on the fact that the spin operator commutes or anti-commutes with $P_z$.
In the case in which the operator commutes with $P_z$ the correlation functions can be evaluated directly. 
Obviously we can not write the explicit form of all the spin correlation functions in terms of fermionic functions. We just limit ourselves to 
describe some characteristic examples as the correlation functions that enter in the reduced density  matrix of two spin at a distance $r=1$.
With some algebra it is easy to show that the expressions of the correlation functions 
in terms of $g(r,t)$, $f(r,t)$ and $h(r,t)$ are the following
\begin{eqnarray}
 \braket{\sigma_i^z}&=&g(0,t)  \\
 \braket{\sigma_i^z\sigma_{i+1}^z}&=&g(0,t)^2-f(-1,t)^2-g(-1,t)g(1,t)\nonumber \\
 \braket{\sigma_i^x\sigma_{i+1}^y}&=&i\,f(-1,t)\nonumber\\
 \braket{\sigma_i^x\sigma_{i+1}^x}&=&g(-1,t) \nonumber\\
 \braket{\sigma_i^y\sigma_{i+1}^y}&=&g(1,t) \nonumber \;.
\end{eqnarray}

Unfortunately, the way to evaluate the correlation functions associated to spin operators that does not commute with the parity operator along
the $z$ spin direction is much more complex and  we have to use the trick that we have discuss in Sec.II. 
For any operator $\hat{O}^{\{\mu_i\}}_S$ that anti-commutes with $P_z$. we define the operator 
$W_{S\cup S+R}^{\{\mu_i\}}=\hat{O}^{\{\mu_i\}}_S \hat{O}^{\{\mu_i\}}_{S+R}$, where $S+R$ is a support obtained from $S$ by a rigid spatial 
translation of $R$ spins and 
$\hat{O}^{\{\mu_i\}}_{S+R}=\sigma_{i_1+R}^{\mu_1}\! \otimes\! \sigma_{i_2+R}^{\mu_2}\! \otimes \!\ldots\!  \otimes \!\sigma_{i_l+R}^{\mu_l} $. 
The operator $W_{S\cup S+R}^{\{\mu_i\}}$, defined on a finite support that is the union of $S$ and $S+R$, commutes with the 
parity operator and hence its expectation values can be evaluated with the standard approach. 
The expectation value of $\hat{O}^{\{\mu_i\}}_S$ is recovered exploiting  the property of asymptotic factorization of products of local operators at 
infinite distance that yields to
\begin{equation}
\label{appendixexpected}
\langle \hat{O}^{\{\mu_i\}}_S \rangle =\sqrt{ \lim_{R \rightarrow \infty}
\langle W_{S\cup S+R}^{\{\mu_i\}} \rangle
}\; ,
\end{equation}
The expectation value in the r.h.s. of eq.~(\ref{appendixexpected}) can be evaluated making use of the Pfaffians that at $t=0$ and $t\rightarrow\infty$
reduce to the standard determinant~\cite{Barouch1970}. 
Usually, with the exception of some particular case at $t=0$ or $t\rightarrow \infty$, it is not possible to evaluate analytically the limit of 
diverging $R$ of the Pfaffians. 
We are forced to make use of numerical evaluation of eq.~(\ref{appendixexpected}). 
Obviously, having to resort to a numerical evaluation, we are forced to limit our analysis to a finite value of $R$, named $R_{max}$,
large but finite. 
We must therefore ask ourselves whether the approximation that we make is valid and if so what are its limits. 
To answer to this question, in fig.~(\ref{nsym_corr_diff}), we reported the difference, evaluated in the $XY$ model, between evaluations made with two 
different $R_{max}$ ($R_{max}$ and $R_{max}+\Delta R$ with $\Delta R$ sets to 10), of four correlation functions that break the parity symmetry. 
If the difference is greater than the computational noise threshold, that we set arbitrary to $10^{-9}$, it means that the estimation of 
$\langle\hat{O}^{\{\mu_i\}}_A\rangle$ made with that $R_{max}$ is not good.  
As we can see in fig.~(\ref{nsym_corr_diff}), all the curves have a very similar pattern. Up to a certain time $ t^*$, that grows with the 
increase of $R_{max}$, the difference between the two values is comparable with the computational noise. 
However, regardless the choice of $R_{max}$, when $t$ becomes greater than $t^*$ one begins to notice a clear, and coherent, increment of the 
difference between the two values which arrives, after a certain transient, to a threshold value that decreases while $R_{max}$ increase. 
However it is nonetheless significant and not negligible with workable value of $R_{max}$. 
For this reason, all the results showed in the main text are obtained considering $t$ always less that $ t^*$.

\end{document}